\documentclass[ aip, jmp, amsmath,amssymb, reprint, ]{revtex4-1}
\usepackage{graphicx}
\usepackage{dcolumn}
\usepackage{bm}
\usepackage{amssymb}
\usepackage{amsmath}
\usepackage{amsfonts}
\usepackage{comment}

\usepackage{hyperref}
\usepackage{bbm}
\usepackage[usenames]{color}
\begin{document}

\preprint{AIP/123-QED}

\title{Nonlocal-coupling-based control of coherence resonance}

\author{Aleksey Ryabov}
\affiliation{Institute of Physics, Saratov State University, Astrakhanskaya str. 83, 410012 Saratov, Russia}

\author{Elena Rybalova}
\affiliation{Institute of Physics, Saratov State University, Astrakhanskaya str. 83, 410012 Saratov, Russia}

\author{Andrei Bukh}
\affiliation{Institute of Physics, Saratov State University, Astrakhanskaya str. 83, 410012 Saratov, Russia}

\author{Tatiana E. Vadivasova}
\affiliation{Institute of Physics, Saratov State University, Astrakhanskaya str. 83, 410012 Saratov, Russia}

\author{Vladimir V. Semenov}
\email{semenov.v.v.ssu@gmail.com}
\affiliation{Institute of Physics, Saratov State University, Astrakhanskaya str. 83, 410012 Saratov, Russia}

\date{\today}

\begin{abstract}
We demonstrate that nonlocal coupling enables control of the collective stochastic dynamics in the regime of coherence resonance. The control scheme based on the nonlocal interaction properties is presented by means of numerical simulation on an example of coupled FitzHugh-Nagumo oscillators. In particular, increasing the coupling radius is shown to enhance or to suppress the effect of coherence resonance which is reflected in the evolution of the dependence of the correlation time and the deviation of interspike intervals on the noise intensity. Nonlocal coupling is considered as an intermediate option between local and global coupling topologies which are also discussed in the context of the coherence resonance control.
\end{abstract}

\maketitle

\begin{quotation}
The phenomenon of coherence resonance is associated with growth of the stochastic oscillation regularity when increasing the noise level in certain range. Therefore, there is an optimal  noise intensity that corresponds to the most regular noise-induced oscillations. Such effects are observed in a broad spectrum of dynamical systems of different nature, which dictates the scientific significance of the issues addressing the coherence resonance control.
Various control schemes can be applied to achieve more or less pronounced coherence resonance as well as to adjust the optimal noise intensity corresponding to the peak regularity of the noise-induced oscillations. The approach discussed in the present paper can be realized in networks of coupled oscillators and implies varying the coupling topology to control the collective dynamics in the regime of coherence resonance. In particular, it is demonstrated that nonlocal coupling provides for enhancing or suppressing coherence resonance when increasing the coupling radius such that transition from local to global coupling is realized. In addition, it allows to shift an optimum intermediate value of the noise intensity in a wide range. Thus, the intrinsic peculiarities of nonlocal coupling make such kind of interactions to be an appropriate tool for coherence resonance control.
\end{quotation}

\section{Introduction}
\label{intro}
A manifold of dynamical systems exhibiting the effect of coherence resonance is incredibly broad where excitable \cite{pikovsky1997,lindner1999,lindner2004,deville2005,muratov2005,semenov2017,semenov2018} and non-excitable oscillators \cite{gang1993,ushakov2005,zakharova2010,zakharova2013,geffert2014,semenov2015} can be distinguished. The stochastic dynamics associated with manifestations of coherence resonance is found to occur in the context of various fields including neurodynamics \cite{pikovsky1997,lee1998,lindner2004,pisarchik2023,tateno2004}, microwave \cite{dmitriev2011} and semiconductor \cite{hizanidis2006,huang2014,shao2018} electronics, optics \cite{dubbeldam1999,giacomelli2000,avila2004,otto2014,arteaga2007,arecchi2009}, quantum physics \cite{kato2021}, thermoacoustics \cite{kabiraj2015}, plasma physics \cite{shaw2015}, hydrodynamics \cite{zhu2019}, climatology \cite{bosio2023} and chemistry \cite{miyakawa2002,beato2008,simakov2013}. The occurrence of coherence resonance can be accompanied by another effects, for instance, synchronization \cite{balanov2009,semenov2025-2} and noise-induced transitions \cite{semenov2017}. In particular, it is known that noise-induced oscillations corresponding to coherence resonance can be synchronized mutually or by external forcing \cite{han1999,ciszak2003,ciszak2004}. Moreover, the synchronization of the noise-induced oscillations occurs in a similar way as for a deterministic quasiperiodic system \cite{astakhov2011}.

A large diversity of dynamical systems exhibiting coherence resonance and related applications necessitates  search for coherence resonance control schemes. Various approaches are applied to regulate the degree of the coherence resonance manifestation or to shift optimal values of noise intensity to a more appropriate range. In particular, introduction of time-delayed feedback allows to control the characteristics of noise-induced oscillations in systems with type-I \cite{aust2010} and type-II \cite{janson2004,brandstetter2010} excitability as well as in non-excitable \cite{geffert2014,semenov2015} coherence resonance oscillators. Alternatively, one can vary the properties of noise to enhance or to suppress coherence resonance. For instance, this can be realized when varying the correlation time of coloured noise \cite{brandstetter2010}. In addition, coherence resonance can be efficiently controlled by varying the L{\'e}vy noise parameters (the noise's probability density function transforms when parameters change) which was demonstrated both numerically and experimentally \cite{semenov2024}. In networks of coupled oscillators, one can vary the coupling strength and modify the coupling topology for controlling coherence resonance, which was successfully demonstrated on examples of multilayer networks with multiplexing \cite{semenova2018,masoliver2021}. In the current paper, we extend a manifold of coupling-topology-based methods by considering the impact of nonlocal coupling on the collective stochastic dynamics in the regime of coherence resonance.

Nonlocal interaction is known to impact various effects including pattern formation associated with the Belousov-Zhabotinsky reaction \cite{hildebrand2001,nicola2006} and Rosensweig instability \cite{friedrichs2003}. Moreover, nonlocal character of spatial interaction can play a significant role in the exhibition of fingering \cite{pismen2017}, remote wave triggering \cite{christoph1999}, Turing structures \cite{li2001}, periodic travelling waves, single and multiple pulses \cite{alfaro2014,volpert2015,achleitner2015}, wave instabilities \cite{nicola2006}, or spatiotemporal chaos \cite{varela2005}. 
The nonlocal diffusion is reported to control the wave propagation in excitable \cite{bachmair2014} and bistable \cite{colet2014,gelens2014,siebert2014,siebert2015} media. In networks of coupled oscillators, nonlocal coupling is responsible for occurrence of chimera \cite{abrams2004,zakharova2020,parastesh2021} and solitary \cite{jaros2018,berner2020,schuellen2022} states, control of stochastic resonance \cite{semenov2025-3} and wavefront propagation \cite{semenov2025-4}. In the present work, a list of phenomena associated with the action of nonlocal coupling is complemented by enhancing and suppressing coherence resonance. To formulate a general conclusion on the impact of nonlocal coupling on  noise-induced resonant phenomena, we carry out a comparative analysis of the obtained results with materials reported in Ref. \cite{semenov2025-3} addressing the nonlocal-coupling-based control of stochastic resonance. 

\section{Model and methods}
\label{model_and_methods}
Systems under current study are ensembles of coupled FitzHugh-Nagumo oscillators generally described by the following equations:
\begin{equation}
\label{eq:general}
\begin{array}{l}
\varepsilon\dfrac{dx_{i}}{dt}=x_i-x_i^3/3-y_i+f_i(x_1,x_2,...,x_N), \\
\dfrac{dy_i}{dt}=x_i+a+\sqrt{2D}n_i(t),
\end{array}
\end{equation}
where $x_i=x_i(t)$ and $y_i=y_i(t)$ are dynamical variables, $i=1,2,...,N$, $N=100$ is the total number of interacting oscillators. A parameter $\varepsilon\ll 1$ is responsible for the time scale separation of fast activator, $x_i$, and slow inhibitor, $y_i$, variables, $a$ is a threshold parameter which determines the system dynamics: the ensemble elements exhibit the excitable regime at $|a|>1$ and the oscillatory one for $|a|<1$. In the present research, each oscillator $x_i$ is considered in the excitable regime ($a=1.05$ and $\varepsilon=0.01$) in the presence of additive white Gaussian noise of intensity $D$ (noise sources $\sqrt{2D}n_i(t)$ are assumed to be statistically independent), i.e., $<n_i(t)>=0$ and $<n_i(t)n_{j}(t)>=\delta_{ij}\delta(t-t')$, $\forall i,j$. Terms $f_i(x_1,x_2,...,x_N)$ are responsible for the action of coupling. Two options for the coupling terms are studied:
\begin{equation}
\label{eq:nonlocal_coupling}
\begin{array}{l}
f_i(x_1,x_2,...,x_N)= \dfrac{\sigma}{2R} \sum\limits_{j=i-R}^{i+R}(x_j-x_i),
\end{array}
\end{equation}
\begin{equation}
\label{eq:global_coupling}
\begin{array}{l}
f_i(x_1,x_2,...,x_N)= \dfrac{\sigma}{N} \sum\limits_{j=1}^{N}(x_j-x_i),
\end{array}
\end{equation}
where $\sigma$ is the coupling strength, $R$ is the coupling radius. Terms (\ref{eq:nonlocal_coupling}) describe the local interaction at $R=1$ and the nonlocal one at $R>1$, whereas Eqs. (\ref{eq:global_coupling}) correspond to global coupling.

Ensemble (\ref{eq:general}) is studied by means of numerical modelling which is carried out by integration of the differential equations using the Heun method \cite{mannella2002} with the time step $\Delta t=0.001$ and the total integration time $t_{\text{total}}=10^3$. The used initial conditions are chosen to be random and uniformly distributed in the ranges  $x_i(t=0)\in[-1.5,1.5]$ and $y_i(t=0)\in[-1.5,1.5]$. Such initial conditions allow to avoid the appearance of travelling waves \cite{korneev2024}, chimera states \cite{semenova2016} and other spatio-temporal structures which can potentially arise in ensemble (\ref{eq:general}) 
 in the presence of weak noise. 

To reveal the intrinsic properties of coherence resonance, we explore the evolution of numerically obtained time realizations $x_i(t)$ caused by the noise intensity growth at fixed coupling parameters. 
This implies visulalization of realizations $x_i(t)$ as space-time plots. After that the correlation time, $t^{\text{cor}}_i$, and the normalized standard deviation of interspike intervals, $R^{ISI}_i$ are calculated for each oscillator: 
\begin{equation}
\label{eq:t_cor_and_R_ISI} 
\begin{array}{l}
t^{\text{cor}}_i=\dfrac{1}{\Psi(0)}\int\limits_{0}^{\infty} \left| \Psi(s) \right|ds, \\
R^{ISI}_i=\dfrac{\sqrt{\left< T_{ISI}^2\right>-\left< T_{ISI}\right>^2}}{\left< T_{ISI}\right>},
\end{array}
\end{equation} 
where $\Psi(s)$ and $\Psi(0)$ are the autocorrelation function and the variance of the time realization $x_i(t)$, $T_{ISI}$ is a sequence of time intervals between neighbour spikes in $x_i(t)$. Finally, the averaged values of the introduced characteristics are extracted to globally characterise the collective dynamics: $\overline{t}_{\text{cor}}=\dfrac{1}{N}\sum\limits_{i=1}^{N}t^{\text{cor}}_i$, $\overline{R}_{ISI}=\dfrac{1}{N}\sum\limits_{i=1}^{N}R^{ISI}_i$. Similarly, the power spectrum averaged over the ensemble, $\overline{S}(\omega)$, is taken into consideration to emphasize the similarity of the effects observed in the ensembles with coherence resonance in a single oscillator: $\overline{S}(\omega)=\dfrac{1}{N}\sum\limits_{i=1}^N S_i(\omega)$, where $S_i(\omega)$ is the power spectrum of the network node oscillations $x_i(t)$.

\section{Local coupling}

\begin{figure}[t!]
\centering
\includegraphics[width=0.49\textwidth]{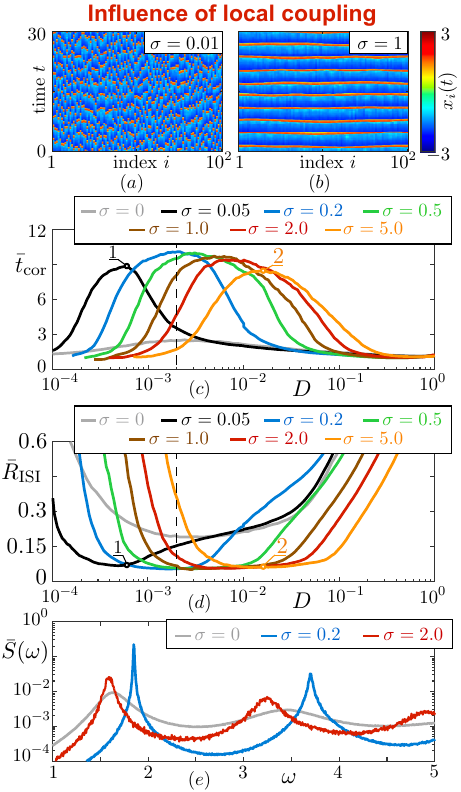}
\caption{Enhancement of coherence resonance in ensemble (\ref{eq:general}) due to the action of local coupling governed by Eqs. (\ref{eq:nonlocal_coupling}) at $R=1$: (a)-(b) Space-time plots illustrating synchronization of noise-induced oscillations when increasing the coupling strength at $D=0.01$; (c)-(d) Dependencies of the averaged correlation time (panel (c)) and normalized standard deviation of interspike intervals (panel (d)) on the noise intensity obtained when increasing the coupling strength; (e) Evolution of the averaged power spectrum of oscillations $x_i(t)$ at fixed noise intensity, $D=0.002$ (corresponds to the vertical dashed line in panels (c)-(d)), and increasing coupling strength. The oscillators' parameters are $a=1.05$, $\varepsilon=0.01$.}
\label{fig1}
\end{figure}

We begin our study from a ring of locally coupled oscillators (see model (\ref{eq:general}) where the coupling term is described by Eqs. (\ref{eq:nonlocal_coupling}) at $R=1$). In the presence of weak coupling the collective stochastic dynamics is asynchronous, the spiking activity of interacting oscillators has spontaneous, independent character which is clearly visible in space-time plots [Fig. \ref{fig1}~(a)]. If the coupling strength is large enough and the noise intensity is chosen such that single oscillators (see Eqs. (\ref{eq:general}) in the absence of the coupling term)  exhibit coherence resonance, one observes synchronous stochastic oscillations  [Fig. \ref{fig1}~(b)]. However, the impact of coupling is not limited by arising the spatial coherence, but also enhances coherence resonance. This effect consists in growth of the local dynamics regularity when increasing the coupling strength, which is illustrated in Fig.~\ref{fig1} on examples of dependencies of the averaged correlation time [Fig.~\ref{fig1}~(c)] and the normalized standard deviation of interspike intervals  [Fig.~\ref{fig1}~(d)] on the noise intensity. Indeed, as demonstrated in Fig.~\ref{fig1}~(c),(d) growth of the coupling strength allows to significantly increase the peak value of $\bar{t}_{\text{cor}}$ and decrease the minimum value of $\bar{R}_{\text{ISI}}$ as compared to the dynamics of the single oscillators. Moreover, increasing the coupling strength provides for  
varying the optimal noise intensity $D_{\text{opt}}$ corresponding to the most coherent oscillations. In particular, the optimal noise intensities at $\sigma=0.05$  and $\sigma=2.0$ are $D_{\text{opt}}\approx 6 \times 10^{-4}$ and $D_{\text{opt}}\approx 6 \times 10^{-3}$ correspondingly, whereas the most regular oscillations of coupling-free ensemble elements are achieved at $D_{\text{opt}}\approx 2 \times 10^{-3}$.

The transformation of curves $\bar{t}_{\text{cor}}(D)$ and $\bar{R}_{\text{ISI}}(D)$ illustrated in Fig.~\ref{fig1}~(c),(d) indicate the option for coherence resonance control at fixed noise intensity. For instance, varying the coupling strength at $D=0.002$ (the vertical dashed line in Fig.~\ref{fig1}~(c),(d)) first enhances coherence resonance. After passing the optimal coupling strength $\sigma_{\text{opt}}=0.2$ corresponding to the most pronounced coherence resonance, the coupling-induced enhancement is changed to suppression of the noise-induced oscillation coherence. The resonant character of the local coupling impact is also reflected in the evolution of the power spectra obtained at $D=0.002$ and increasing $\sigma$ [Fig.~\ref{fig1}~(e)]. It is clearly visible that the maximal values of $\bar{t}_{\text{cor}}$ and minimal $\bar{R}_{\text{ISI}}$ achieved at $\sigma = 0.2$ correspond to the most prominent main peak in the averaged power spectra of oscillations $x_i(t)$. Thus, the evolution of quantities $\bar{t}_{\text{cor}}$ and $\bar{R}_{\text{ISI}}$ is in complete accordance with the transformations of power spectra. 

\begin{figure}[t!]
\centering
\includegraphics[width=0.49\textwidth]{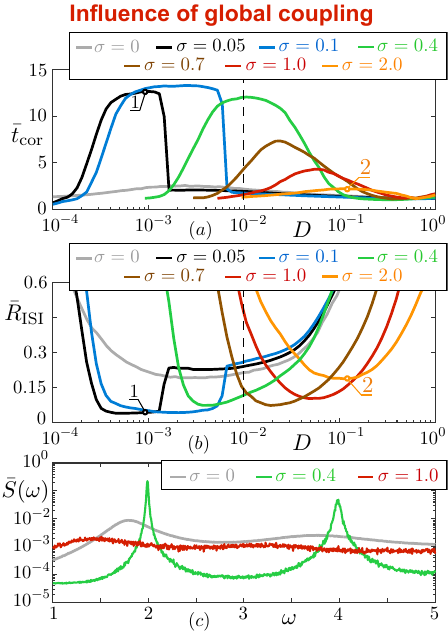}
\caption{Enhancement of coherence resonance in ensemble (\ref{eq:general}) due to the action of global coupling described by Eqs. (\ref{eq:global_coupling}): (a)-(b) Dependencies of the averaged correlation time (panel (a)) and normalized standard deviation of interspike intervals (panel (b)) on the noise intensity obtained when increasing the coupling strength; (c) Evolution of the averaged power spectrum of oscillations $x_i(t)$ at fixed noise intensity, $D=0.01$ (corresponds to the vertical dashed line in panels (a)-(b)), and increasing coupling strength. The oscillators' parameters are $a=1.05$, $\varepsilon=0.01$.}
\label{fig2}
\end{figure}

\section{Global coupling}
Similarly to the influence of local interaction, global coupling also provides for enhancement of coherence resonance. As depicted in Fig. \ref{fig2}~(a),(b) on an example of ensemble (\ref{eq:general}) where the coupling terms are described by Eqs. (\ref{eq:global_coupling}), global coupling enables to increase the peak values of the correlation time and decrease the minimal values of $\bar{R}_{\text{ISI}}$. In addition, global kind of interaction allows to increase the optimal value of the noise intensity, $D_{\text{opt}}$, corresponding to the most coherent oscillations. Shift of the optimal noise intensity results in the resonant enhancement of coherence resonance when increasing coupling strength at fixed noise intensity, which complements the similarity with the action of local coupling (compare the power spectrum evolution in Fig.~\ref{fig1}~(e) and Fig.~\ref{fig2}~(c)). However, there is the principal difference between the impact of local and global coupling manifested at high values of the coupling strength. In contrast to the effects exhibited in the presence of local coupling, the global-coupling-induced enhancement of coherence resonance has a resonant character. Indeed, growth of the coupling strength first increases the oscillation regularity, but then the oscillations become less and less coherent. Resultantly, the curves $\bar{t}_{\text{cor}}(D)$ and $\bar{R}_{\text{ISI}}(D)$ obtained in the absence of coupling and in the presence of strong interaction are not much different in the context of enhancement or suppression of coherence resonance: the only difference taking place is a horizontal shift (compare the grey and orange curves in Fig. \ref{fig2}~(a),(b) obtained at $\sigma=0$ and $\sigma=2$). As demonstrated in the next section, this established aspect can play a key role and results in coherence resonance control when increasing the radius of nonlocal coupling. 

\section{Nonlocal coupling}
\label{sec_nonlocal}
The comparative analysis of dependencies in Fig.~\ref{fig1}~(c),(d) and Fig.~\ref{fig2}~(a),(b) allows to conclude that in case of low coupling strength the global topology is more efficient in the context of the stochastic resonance enhancement as compared to the local interaction. In particular, one can compare the black curves corresponding to $\sigma=0.05$: the peak value of $\bar{t}_{\text{cor}}(D)$ registered at the optimal noise intensity in the presence of global coupling is higher than the one obtained in case of local coupling ($\bar{t}_{\text{cor}}^{\text{peak}}=12.6$ versus $\bar{t}_{\text{cor}}^{\text{peak}}=9.87$, see points 1 in Fig.~\ref{fig1}~(c) and Fig.~\ref{fig2}~(a)). Similarly, the corresponding value of $\bar{R}_{\text{ISI}}$ is lower in the case of global coupling ($\bar{R}_{\text{ISI}}^{\text{min}}=0.0385$ versus $\bar{R}_{\text{ISI}}^{\text{min}}=0.0674$, see points 1 in Fig.~\ref{fig1}~(d) and Fig.~\ref{fig2}~(b)). In contrast, local coupling enhances coherence resonance more efficiently when the coupling strength is high (compare the values of $\bar{t}_{\text{cor}}$ and $\bar{R}_{\text{ISI}}$ at points 2 in Fig. ~\ref{fig1}~(c),(d) and Fig. ~\ref{fig2}~(a),(b)). Thus, either local or global coupling topologies is more appropriate for enhancement of coherence resonance at different coupling strengths. In such a case, one can expect that increasing the radius of nonlocal coupling as a continuous transition from local to global coupling could enhance or suppress the stochastic resonance which depends on the coupling strength. To prove this hypothesis, further study of ensemble (\ref{eq:general}) is carried out in the presence of coupling terms (\ref{eq:nonlocal_coupling}) where $R$ increases at fixed $\sigma$.

\begin{figure}[t!]
\centering
\includegraphics[width=0.49\textwidth]{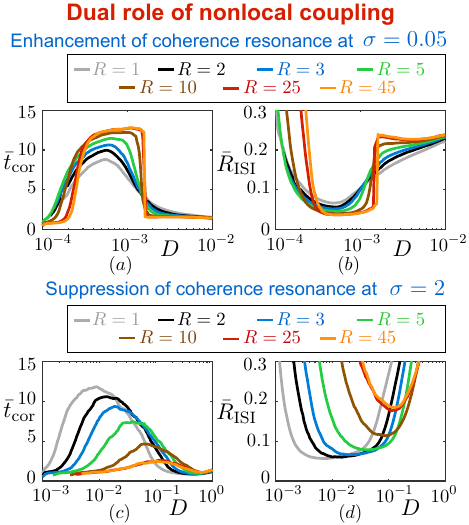}
\caption{Enhancement (panels (a) and (b)) and suppression (panels (c) and (d)) of coherence resonance when increasing the coupling radius in model (\ref{eq:general}) with nonlocal coupling (\ref{eq:nonlocal_coupling}). The evolution of the dependencies of the averaged correlation time (left panels) and normalised deviation of interspike intervals (right panels) on the noise intensity are used to visualise the coherence resonance control. The oscillators' parameters are $a=1.05$, $\varepsilon=0.01$. The coupling strength is $\sigma=0.05$ (panels (a) and (b)) and $\sigma=0.05$ (panels (c) and (d)).}
\label{fig3}
\end{figure}

Character of the stochastic dynamics evolution caused by increasing the coupling radius is illustrated in Fig.~\ref{fig3} for two options: $\sigma=0.05$ and $\sigma=2$ (panels (c) and (d)). In the first case, increasing the coupling radius enhances coherence resonance [Fig.~\ref{fig3}~(a),(b)]. The effect of saturation is observed at $R>25$ and further growth of the coupling radius does not lead to noticeable changes of the dynamics. In contrast, growth of the coupling radius suppresses coherence resonance at $\sigma=2$ [Fig. \ref{fig3} (c),(d)]. As emphasized above, such value of $\sigma$ corresponds to less pronounced global-coupling-induced enhancement of stochastic resonance, whereas the local coupling provides for much more effective supporting the stochastic oscillation regularity. As a result, increasing the coupling radius allows to realize continuous transition between local and global coupling and to observe the evolution of the curves $\bar{t}_{\text{cor}}(D)$ and $\bar{R}_{\text{ISI}}(D)$ such that coherence resonance is suppressed, which also has a saturable character: no visible changes of the dynamics are observed when the coupling radius exceeds the value $R=25$.

A distinguishable effect is observed at the low coupling strength and large enough coupling radius: the second local peak in the dependency $\bar{t}_{\text{cor}}(D)$ [Fig.~\ref{fig3}~(a)] and the second local minimum in the curve $\bar{R}_{\text{ISI}}(D)$ [Fig.~\ref{fig3}~(b)] appear. The same effect takes place in the system with global coupling (see the black curves in Fig.~\ref{fig2}~(a),(b) obtained at $\sigma=0.05$). Despite the second extrema in the dependencies are much less pronounced, such stochastic effect can be referred to the double-peak coherence resonance \cite{sethia2007}.

\section{Conclusions}
Coupling excitable oscillators represents a powerful tool for increasing the regularity of noise-induced oscillations in the regime of coherence resonance as compared to the coupling-free dynamics. 
The enhancement of noise-induced coherence is manifested as increasing the peak values of the correlation time and decreasing the minimal values of the interspike interval deviation which completely correlates with 
the evolution of the power spectra. To achieve the most coherent oscillations, one can apply different coupling topologies. In particular, either global or local coupling is more efficient in the context of coherence resonance enhancement at different coupling strengths. Is such a case, transition between two topologies as increasing the radius of the nonlocal interaction provides for enhancing (at low coupling strength) or suppressing (in the presence of strong coupling) coherence resonance. These effects are reflected in increasing or decreasing the peak values of the correlation time and the minimal values of the deviation of interspike intervals when varying the coupling radius. Thus, using the nonlocal coupling appears as a promising strategy for controlling stochastic resonance.

Intriguingly, all the aspects of the impact of local, nonlocal and global coupling on coherence resonance mentioned above are very similar to the influence of the mentioned kinds of coupling on stochastic resonance \cite{semenov2025-3}. This fact indicates the fundamental character of the effect of coupling on noise-induced resonant phenomena. In addition, the obtained results are in a good agreement with materials reported in  paper \cite{bukh2025} where the impact of nonlocal interactions on the dynamics of excitable oscillators is discussed in the context of small-size networks. Note that the presented results also correlate with conclusions of article \cite{masoliver2017} where the problem of coherence resonance control in ensembles of coupled excitable oscillators is considered on an example of nonlocal time-delayed coupling. 

The findings of the present work are the basis for further studies such as analysis of the appearance of the second local minima and maxima in the dependencies of the correlation time and the interspike interval deviation on the noise intensity. In addition, revealing the theoretical reasons for the coherence resonance control by means of analytical approaches is a logical continuation of the current research. Furthermore, character of the coupling impact on the resonant noise-induced phenomena can change when varying the system size, which was demonstrated on examples of stochastic \cite{pikovsky2002} and coherence \cite{toral2003} resonances in ensembles of globally-coupled oscillators. Thus, similar effects are expected to occur with an increase in the coupling radius in ensembles of nonlocally coupled oscillators of different size. However, this problem requires a separate consideration.

\section*{DATA AVAILABILITY}
The data that support the findings of this study are available from the corresponding author upon reasonable request.

\section*{Acknowledgements}
This work was supported by the Russian Science Foundation (project No. 23-72-10040). 

\text{https://rscf.ru/project/23-72-10040/}



%

\end{document}